   \documentclass[12pt]{article}
   \usepackage{amssymb}
   \usepackage{amsmath}
   \usepackage{epsfig}
   \newcommand{\be}{\begin{equation}}
   \newcommand{\ee}{\end{equation}}
   \newcommand{\bea}{\begin{eqnarray}}
   \newcommand{\eea}{\end{eqnarray}}
   \makeatother
   \makeatletter
   \renewcommand{\@biblabel}[1]{#1.}
   \makeatother \textheight 24 cm \voffset -3 cm
   
\begin{document}
   \title{ Tuned electron-nucleus resonance as a tool of producing the $^{229m}$Th isomer }
   \author{F. F. Karpeshin }
   
   \maketitle
   \begin{center}
   D. I. Mendeleyev Institute for Metrology, Saint-Petersburg, Russia
    \\ e-mail {\it fkarpeshin@gmail.com}
   \end{center}
   \begin{abstract}
   The possibility of refining the energy of the 8.36-eV $^{229m}$Th nuclear isomer --- the most likely candidate for the role of a nuclear frequency standard --- by means of resonant optical pumping is discussed. Attention is focused on considering the problem of broadening the resonance in order to reduce scanning time. The two-photon method proposed exploits the radical broadening of the isomer line due to mixing with the electron transition. This is not burdened with the cross-section reduction, in contrast with internal-conversion-based resonance broadening  or intended extra-broadening of the spectral line of a pumping laser. In the case under consideration, according to the calculations, it turns out to be two orders of magnitude more efficient. It is applicable to both ionized and neutral thorium atoms. Realization of the method supposes excitation of the both nucleus and electron shell in the final state. 
   \end{abstract}

   \large
   \bigskip
   
   \newpage
   \section{Introduction}

      Much attention is being paid to the problem of creating nuclear optical clocks and, accordingly, the next generation frequency standard. Record samples of atomic clocks demonstrate an error within several units of $10^{-18}$ \cite{pik}, while in order to solve challenging fundamental and applied problems it is necessary to further reduce the errors by another order of magnitude. The development of heavy-ion clocks \cite{kozclock} has good prospects. A further reduction in the error would allow to resolve the long-standing question about the possible drift of the fundamental constants \cite{dreif}. The most pressing task of modern physics is the search for dark matter and energy. Here the fundamental idea is to detect wave oscillations of particles of ultralight matter in its interaction with ordinary matter \cite{snowmass}. The strengths of nuclear clocks are indicated in many reviews and original works (for example, \cite{randevu} and references therein). Some of them seem self-evident, such as protection from external fields. And their use to search for the drift of fundamental constants has irreplaceable features, since the contribution from the nuclear component, compared to the Coulomb component, to the transition frequency is much stronger than in optical ones. Some projects are based on the joint use of atomic and nuclear clocks, using the specified features of the latters \cite{kozclock,inter}.
      
      The number one candidate for the creation of nuclear  clocks is the unique nuclide of $^{229}$Th, whose excited state $3/2^+[631]$ lies at a height of only $\omega_n$ = 8.355740(3) eV above the ground state $5/ 2^+[633]$ \cite{BCSH}. For estimates, I set its lifetime in neutral atoms to be within 10 $\mu$s \cite{lars1,prc2007,isakov,zhang}. The proper lifetime of the nucleus is much longer due to internal conversion (IC), which greatly enhanced the decay rate of the isomer: ICC (internal conversion coefficient) $\alpha(M1) = 0.987\times10^9$. Thus,  accounting for the IC leads to an increase in the natural width of the isomeric line from $\Gamma_n = 0.667\times10^{-19}$ eV ($10^{-5}$ Hz) to $\Gamma_a = 0.7\times10^{-10}$ eV ( 10 kHz). An indication has been obtained that in a crystalline environment it can be several times shorter than \cite{BCSH,sandro}, this not changing the qualitative conclusions, though.
   
      The recent breakthrough \cite{BCSH} was achieved using a highly doped CaF$_2$ crystal and a four-wave mixing laser. As a result, the thorium isomer transition was optically excited resonantly for the first time. Theory of direct nuclear excitation in strong electromagnetic fields  was advanced in Refs. \cite{palfi1,palfi2,dzub1,dzub2,wense} and  cited therein. In this concern, I remind that another effective way to search for nuclear optical resonance is offered by exploiting the resonant properties of the electron shell. The interaction of the nucleus with the electron shell is carried out through internal conversion (IC), which in the subthreshold region turns into discrete, or resonance conversion (RC). The RC concept was formulated yet in relation to deexcitation of fission fragments in muonic atoms \cite{DF,book}. RC was confirmed experimentally in Ref. \cite{david}. A method for laser-induced radiative decay of the 76-eV isomer $^{235}$U by means of RC was proposed in Ref. \cite{kabzon}. The method is based on two-photon resonance absorption by the atomic electrons: one photon --- virtual --- is emitted by the nucleus, the other comes from the external radiation of a tunable laser, which is used in order to tune the resonance. In this way,  the question of scanning the laser frequency range at the isomer energy was first posed. Common features of RC with the 
electronic bridges (EB), introduced by Krutov \cite{krutov}, were noted.
It is worth noting that EB was observed experimentally  in  \cite{logan} in the decay of the 30.7-keV isomeric level $^{93}$Nb. RC is actually the cross-invariant continuation of EB into the subthreshold field. In the form of bound internal conversion (BIC) it was discovered in the 35-keV transition in highly charged $^{125}$Te ions \cite{atalah}. 
      
      Based on Ref. \cite{morita}, application of inverse EB was proposed in Ref. \cite{tkalya}. Morita suggested that creation of  holes in inner shells can induce a non-radiative electron transition accompanied with excitation of the nucleus (known as NEET).  Actually guided by the misunderstood paper by I. S. Batkin \cite{batkin}, Tkalya suggested that lifting up a valence electron to an excited level may induce excitation of the nucleus.
      For the nucleus to be effectively excited, however, the intermediate electron state should have exactly the same energy as  the nuclear transition energy.  In fact, Batkin  never told that such a resonance can be realized in nature indeed,  considering energy transfer through a virtual electron level. 
      
      In view of the improbability of such a coincidence, as well as the uncertainty of the isomer 
energy,  a ``non-resonance'' method was proposed in Ref. \cite{PL1}. According to the method,  resonating electronic levels are populated in a spontaneous transition from a higher shell. Moreover, it was shown that in fact, there is a number of the levels which {\it form} resonances with the isomer transition. As a result, {\it all} these resonance intermediate states are populated automatically, contrary to the method of Ref. \cite{tkalya}. An experiment aimed at testing this mechanism was carried out in Ref.  \cite{tom}. The problem, however, was that the most probable isomer energy at that time was considered to be 3.5$\pm$1 eV. The value of 7.6$\pm$1 eV \cite{beck}, close to the present one,  appeared much later.
      
      In Ref. \cite{tai} the authors realized that existence of a resonant pair of levels is hardly possible, supposing that the resonance condition may be not so important if the intermediate electron state is in the continuum. Numerical estimations were performed in Ref. \cite{dzub}. Furthermore, when it became clear that the energy of the isomer lies in the region of 7 -- 8 eV, this induced appearance of projects based on the absorption of two or three laser photons (\cite{porsev,prc2f,muler,Th35,2f,china,2fchina} and Refs. cited therein). At the same time, single-photon projects are also considered, like Ref. \cite{prev}, which became the prototype of the state-of-the-art project \cite{zhang}.

      Whatever project we take, however, we will definitely be faced with the issue of optimizing the scanning step by spectral broadening the line of the pumping beam. Thus, in Ref. \cite{china} it was noted  that the effective width of the isomer line is 40 nuclear widths, which, however, is still too small to make it as a scanning step. In the \cite{zhang} project, the line broadening due to the IC reaches 9 orders of magnitude, but exactly the same amount is lost in the excitation cross section. In the latest paper Ref. \cite{BCSH} the laser had an inherent bandwidth of about 1 GHz, but it was deliberately modulated to an effective bandwidth of 10-20 GHz to have a broader lineshape available for better coverage of the search region. As a consequence, the resonance excitation cross section of the isomer again decreases by the same factor. Different physics lies in the idea of excitation of the nucleus through electron-nuclear resonance. For example, the natural width of the $7p$ level is eleven orders of magnitude greater than the nuclear one, whilst its excitation cross section does not suffer at all, but on the contrary, gains many orders of magnitude in comparison to the nuclear cross section. The combination of these factors creates  attractive prospects for scanning. In order to study them in more detail, I conduct a comparative analysis of the above-mentioned methods, in order to outline ways to further improve the efficiency of research. My narration comes as follows. 
      In the next section I remind the state-of-the-art  project in more details in order to highlight the principled points by the comparison. Section \ref{method} outlines the physical principles of the present method. Neutral atoms are considered. Values needed for assessment of effectiveness of the method   are presented in  section \ref{calculation}. Required  scan time is assessed in section \ref{scantime}. The main results obtained are discussed in section \ref{conclusion}. Prospects for further research are also outlined there.

      \section{ Application of internal conversion in the frequency comb method }
      \label{zhang}

      Start with the state-of-the-art project Ref. \cite{zhang}. Two decisive factors underlie the project.  First, it is frequency comb consisting of  $10^5$ teeth. Broadening of the isomer line by a factor of $10^9$ due to deexcitation via IC comprises the  second factor, contracting  the necessary  scanning time. 

  	Project \cite{zhang} is based on making use of the seventh harmonic of the reference beam from an ytterbium-doped fiber laser with a wavelength of 1070 nm .  Pulse trains from that setup are repeated with a frequency of 77 MHz. As a result, the Fourier spectrum of the transformed beam takes the form of a frequency comb around the target line, consisting of $1.2\times10^5$ equidistant teeth. The radiation power in  each of the teeth  is 10 nW in the center, the half-width is 490 Hz ($2 \times10^{-12}$ eV), the distance between the teeth is 77 MHz = $3 \times 10^{-8}$ eV. This  project was calculated at a time when the isomer energy was considered to be $\omega_n$ = 8.338 eV with an uncertainty of 0.024 eV, according to \cite{sandro}. Thus, the entire comb covers all the uncertainty interval of the isomer energy.

The  project was calculated at a time when the isomer energy was considered to be $\omega_n$ = 8.338 eV with an uncertainty of 0.024 eV, according to \cite{sandro}. Thus, the entire comb was designed to cover the entire uncertainty interval of the isomer energy, and to find the resonance it was necessary to scan only the interval between the teeth.

Furthermore, each tooth of the comb carries a photon flux of 
$J= 1.06\times10^{13}\text{ c}^{-1}\text{cm}^{-2}$.
This flux is focused onto a target, which is a thin circle with an area of 
$7.07\times10^{-4} \text{ cm}^2$. $1.6\times10^{13}$ atoms of $^{229}$Th are deposited onto its surface by sputtering or otherwise. If the isomer energy coincides with a photon one, the nucleus transfers to the isomeric state. The latter decays back through  IC. Detection of the conversion electrons shows that the resonance is established. 

      The cross-sections of photon absorption are related to the radiative widths of the reverse transitions by the detailed balancing principle. For the scanning purposes, it is most effective to use a light beam with a spectral width which is within the resonance width. Then the photoexcitation cross section for a level with energy $\omega$ (we use the relativistic system of units $\hbar = c = m_e$ = 1) can be estimated by means of  the following interrelation \cite{npa,beres}:
      \be
      \sigma_\gamma = \Gamma_\gamma\frac{2 I_\text{is}+1}{2I_0+1}
      \left( \frac{\pi}{\omega}\right)^2S_\omega\,,  	\label{sig}
      \ee    
      where $S_\omega$ is the spectral density of the beam,  $\Gamma_\gamma$ --- the radiative width of the reverse transition. Designating $S_\omega\approx 1/\Gamma$,  $\Gamma$ being the total width of the isomeric state, and supposing, where this is possible, $\Gamma_\gamma\approx \Gamma$, one arrives at the universal expression as follows:
      \be
      \sigma_\gamma = \frac{2I_\text{is} +1}{2I_0+1}
      \left( \frac{\pi}{\omega}\right)^2\,,  		\label{meansig}
      \ee   
      Eqs. (\ref{sig}) -- (\ref{meansig}) are equally valid for both nuclear and atomic systems. Such a universality is extremely important when considering the phenomena of electron-nuclear resonance. In particular, Eq. (\ref{meansig}) describes the cross-section for resonance scattering on a bare nucleus.
      
      In the case of a neutral thorium atom, however, $S_\omega = 1/\Gamma_c$, where $\Gamma_c$ is the conversion width of the isomeric state. Substituting it  in formula (\ref{sig}), one arrives at the expression for the excitation cross-section of the isomer in neutral atoms as follows:
      \bea
      \sigma_\gamma = 
      \frac{\Gamma_\gamma}{\Gamma_c}\ \frac{2I_\text{is}+1}{2I_0+1}
      \left( \frac{\pi}{\omega}\right)^2   = 
      \alpha^{-1}(M1)   \frac{2I_\text{is}+1}{2I_0+1}
      \left(\frac{\pi}{\omega}\right)^2 = 
      3.73\times10^{-20} \text{ cm}^2  \,.  
      \label{sigc}     \eea   
      
      Now one can estimate the rate of isomer formation $p = \sigma_\gamma J = 3.95\times10^{-7}\text{ c}^{-1}$. It is significantly lower than the rate of their spontaneous deexcitation $\gamma=10^5\text{ c}^{-1}$. As a result, $N^\text{(imp)} = \sigma_\gamma JN_0\tau_\text{imp} = 633$ isomeric nuclei are formed per pulse, 90\% of which, however, decay by the beginning of the measurement,  about 60 of them surviving.
      Furthermore, if the laser frequency is changed each second with a step equal to the width of the target line, that is, $10^{-10}$ eV, then scanning of the supposed uncertainty interval of 0.024 eV will require 5000 steps, which takes 5000 seconds. To determine which tooth resonates, one needs to change the interval between the teeth. Therefore, to clarify the energy of the isomer, several more scanning cycles will be required.

   The above project teaches a lesson how the acceleration of the isomeric transition, arising due to IC, can be used in order to reduce the scan time.  Strictly speaking, this can be called a kinematic gain as  no resonance  is exploited. And this way is only suitable for neutral atoms. Already in singly-charged ions, the ionization energy becomes greater than the isomer energy,  turning off the IC channel. At the same time, most clock projects are aimed at  usage of $^{229}$Th ions. Let us see how the Warsaw effect in combination with resonance can be exploited both in neutral and ionized atoms.
   
   \section{Physical premises of application of the resonantly enhanced Warsaw effect}
   \label{method}

      Quite recently, another way of measuring the energy of an isomer was proposed in Ref. 
\cite{Letter}, based on the Warsaw effect of mixing nuclear states with different spins through interaction with an orbital electron. It descends from the fact that the excitation of the isomer occurs through the sudden removal of one of the $s$-shell electrons, according to the Feinberg---Migdal theory \cite{land}. Therefore, in the case of the photoelectric effect, the ejected electrons have  different energies depending on whether the nucleus remains in the ground or isomeric state. By measuring this difference, one obtains the energy of the isomer. Let us see how electron transfer to a discrete $p$-level can  be used to accurately determine the isomer energy. Some directions for the development of these ideas were discussed at the KVNO-23 conference and in its Proceedings \cite{Arxiv}.

   First, I recall the results of Ref. \cite{Zyl},  where  acceleration of the  isomer decay in hydrogen-like ions was demonstrated, occurring due to mixing of the ground and isomeric levels of the nucleus in atomic states with a total angular momentum $F$ = 2. In the ground state of such a hydrogen-like system the spins of the electron and the nucleus can be parallel or antiparallel. The corresponding total moment will be equal to $F$ = 3 or 2. The spins of the electron and the isomeric nucleus are added in a similar way, forming states with $F$ = 2 or 1. Thus, the state of an atom with $F$ = 2 can be of two types: 1) the nucleus is in the ground state, the spins are antiparallel, and 2) the nucleus is in the isomeric state, the spins are parallel. And since they interact through RC, then, according to the principles of quantum mechanics, these states form a superposition, which will be the real wave function. Denoting the amplitudes of mixing of atomic states of the two types as $\alpha$ and $\beta$, the wave function of the atom in the ground or isomeric states can be written as follows:
   \bea
   \Psi_g = \alpha|g\rangle +\beta|is\rangle  \nonumber \\
   \Psi_\text{is} = \alpha |is\rangle - \beta |g\rangle\,,  \label{mx}
   \eea
where $\alpha \approx$ 1 and $\beta$ can be determined through the interaction matrix element and the energy difference between two states in the first order of perturbation theory. And since the interaction is nothing more than a conversion amplitude, the probability of mixing in Eq. (\ref{mx}) can be expressed in terms of radiative nuclear width times ICC as the product of the radiative width of the nuclear transition and discrete conversion coefficient $\alpha_d(M1;1s-1s)$ \cite{Zyl}):
   \be
   \beta^2=\frac{2I_\text{is}+1}{2I_0+1}\frac{\alpha_d(1s-1s;\omega_n)  
   \Gamma_n^{(\gamma)}/2\pi} 
   {\omega_n^2}  \,. \label{gai}
   \ee 
   Here $\omega_n$ is the isomer energy. $\Gamma_n^{(\gamma)}$ --- the radiative partial nuclear width, $\Gamma_n^{(\gamma)} \equiv \Gamma_n$ in the present  case.  
$\alpha_d(1s-1s;\omega_n)$ --- dimensional IC  coefficient (ICC) for the $M1$ nuclear transition with energy $\omega_n$,  and what is more, in this case  a diagonal transition between $1s$-states 
\cite{DF,book,atalah} takes place. It can be concluded that if such an atom is in a state with total momentum $F=2$, then in its ground state the nucleus can nevertheless be in the isomeric state with the probability $|\beta|^2$, and {\it vice versa}, if the atom is in the isomeric state, however, the nucleus can be in the ground one with the same mix  probability $|\beta|^2$.
        Now imagine that one suddenly removes a $7s$ electron from a neutral thorium atom. Then the interaction leading to the formation of an isomer is switched on at the same moment. According to the Feinberg---Migdal theory, the nucleus will remain in the isomeric state with the probability of $\beta^2$.
   
\begin{figure}
\includegraphics[width=0.7\textwidth]{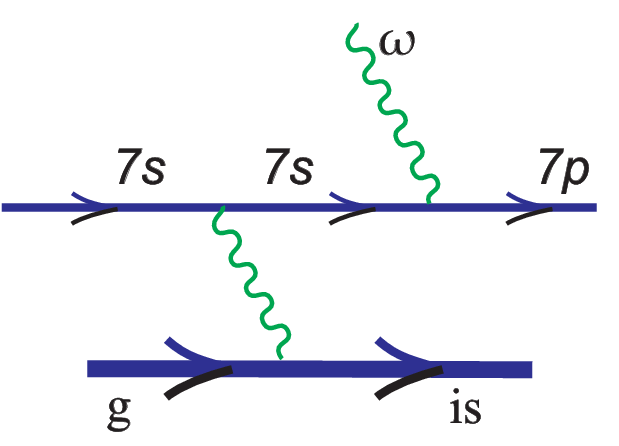}
\caption{\footnotesize Feynman graph of the resonance optical pumping the isomer. The thicker line specifies the nuclear transition from the ground to the isomeric state.}
\label{f2}
\end{figure}
      The Feynman graph of this process is shown in Fig. \ref{f2} with the only difference that the electron is transferred to the $7p$ level. Concerning  the excitation of the isomer, such a transfer has essentially  the same effect, since in the $7p$ state the electron's inelastic interaction with the nucleus can be neglected. In more detail, the nucleus spontaneously transfers into the isomeric state as a result of virtual exchange with a photon with a valence $7s$ electron. The electron falls to $\omega_n$ energy level below the mass shell. The energy balance is restored after it absorbs a photon from the laser field and settles in the $7p$ state of the atom.
The photon energy $\omega$ is determined from the energy conservation: 
\be
\omega = \omega_n + \epsilon_{7p}   \,,    \label{en}
\ee
where $\epsilon_{7p}$ is the energy of the $7p$ level.  At this frequency, the amplitude of the process passes through a resonance with a width $\Gamma_a=\Gamma_{7p}$, determined by the width of the $7p$ level. 

      To calculate the cross section of the process in Fig. \ref{f2}, one can apply formula (\ref{sig})  for the resonance scattering of photons by the $7s$-electron with its transfer to the $7p$-state, provided that the nucleus simultaneously passes into the isomeric state. Since these two processes are independent, one arrives at the following expression for the  cross section:
\be
\sigma_\gamma = \beta^2\frac{2j_{7p}+1}{2j_{7s}+1}
\frac{\Gamma_\gamma^{(a)}(7p-7s; \omega)} {\Gamma_{7p}}
\left(\frac{\pi}{\omega}\right)^2 \,, \label{si}
\ee    
with $\Gamma_\gamma^{(a)}(7p-7s; \omega)$ being the radiative width of the $7p-7s$ transition provided its energy $\omega$ is out of the mass shell, and $\Gamma_{7p}$ --- the total width of the $7p$ level. In view of that the radiative width is proportional to the energy in cube, one can put down 
$\Gamma_\gamma^{(a)}(7p-7s; \omega)= f \Gamma_\gamma^{(a)}(7p-7s; \omega_a)= 
f \Gamma_{7p}$, $f=(\omega/\omega_a)^3$, $\omega_a$ being the energy of the $7p-7s$ transition in a sole atom. By means of these expressions, the cross-section acquires the following form:
\be
\sigma_\gamma = \frac{(2j_{7p}+1)(2I_{is}+1)}{(2j_{7s}+1)(2I_g+1)}
\left(\frac{\pi}{\omega}\right)^2fWQ \,, \label{si}
\ee  
where $Q\leq 1$ is the ratio of the partial radiative width to the total width. This value characterizes the quality factor of the atomic resonance. In our case, $Q$ = 1 --- the maximum value. $W\equiv \beta^2$ is the value of the Warsaw effect (\ref{gai}), in this 
case --- on the $7s$ electrons instead of $1s$. The cross section (\ref{si}) can be transparently represented as follows:
\be
\sigma_\gamma  = \frac{2j_{7p}+1}{2j_{7s}+1}\frac{2I_0+1}{2I_{is}+1}
fWQ \tilde\sigma_\gamma^{(b.n.)}\,,  \label{sigres}
\ee
where $\tilde\sigma_\gamma^{(b.n.)}$ is approximately equal to the cross section on a bare nucleus, differing from the latter in that it is calculated for the photon energy $\omega$ instead of $\omega_n$. It follows from Eq. (\ref{sigres}) that it is reasonable to choose wide levels with good quality factor, for which $Q\approx 1$, as the final state. The same formula  can be obtained in another way, by substituting into Eq. (\ref{sigres}) the radiative width of the process inverse to Fig. \ref{f2}, which was calculated in Ref. \cite{npa}.

      Note that the full resonance width is in the denominator of the formula (\ref{si}), as well as in Eq. (\ref{sig}).  It  does not differ much from the conversion width of the isomer. At the same time, tremendous amplification comes from the vertex corresponding to the electric dipole absorption of a laser photon by the atomic electron. It is by 12 orders of magnitude as large as the magneto-dipole nuclear vertex corresponding to the direct absorption of the photon by the nucleus. 
      
      Finally, for the purposes of qualitative analysis, I introduce the effective cross-section parameter $V$, multiplying the  cross-section (\ref{sigres}) in my case, and (\ref{sigc}) --- in the case of the  project \cite{zhang} by the width of the corresponding resonance, which can be identified with the scanning step, and dividing by the isomer's own width $\Gamma_n$. In the latter case, I arrive at the following transparent expression: 
\be
V \equiv \sigma_\gamma^{(b.n.)}\,,  	\label{bn}
\ee 
In the case of shake-up mechanism in Fig. \ref{f2}, the width is $\Gamma_{7p}$, consequently,
\be
V = fWQ\frac{\Gamma_{7p}}{\Gamma_n} \tilde\sigma_\gamma^{(b.n.)}\,.   \label{bn1}
\ee
In the case of a scheme similar to Ref. \cite{BCSH}, based on intended spectral broadening of the laser beam, Eq. (\ref{bn}) holds for the effective cross section.

\section{ Calculation results }
	\label{calculation}
      
        For the purpose of a qualitative assessment of the scanning time using the mechanism in Fig. \ref{f2}, one can use the Dirac---Fock method. Within this method, required  parameters were calculated using the computer code RAINE \cite{RAINE}. They are published e.g. in Ref. \cite{npa} and elsewhere. Those calculations had been performed using the isomer energy of 3.5 eV. Extrapolation to modern energy is performed directly, using the energy dependence of the radiative widths $\sim\omega^3$ and ICC as $\sim\omega^{-3}$. The obtained values were published in a number of works, for example, in \cite{npa}. Extrapolation to modern energy is performed directly using the energy dependence of radiation widths $\sim\omega^3$ and conversion coefficients $\sim\omega^{-3}$.
      As a result, ICC for the $M1$ transition with energy $\omega_n$ = 8.36 eV in Th I turned out to be equal to $\alpha(M1) = 0.987\times10^9$ \cite{prc2007}. With  the lifetime in neutral atoms of 10 $\mu$s, the above-mentioned proper width of the isomer $\Gamma_n = 0.667\times10^{-19}$ eV follows. 
      
         Furthermore, within the framework of the method, the resonance cross-section consists of the two components corresponding to the final levels either $7p_{1/2}$ or $7p_{3/2}$, the intensity of the second component being twice as high as the first one. Energy of the final state $\epsilon_{7p_{3/2}}$ = 1.64 eV \cite{npa}, therefore $\omega$ = 10.75 eV. In turn, $\epsilon_{7p_{1/2}}$ = 1.64 eV, respectively $\omega$ = 9.98 eV for this level. I also  get $\alpha_d(7s-7s;\omega_n) = 1.23\times10^{10}$ eV.
      Radiative width of an atomic transition
      $\Gamma_\gamma^{(a)}(7p_{3/2}-7s;\omega) = 3.1\times10^{-6}$ eV, which is thirteen orders of magnitude greater than the natural width of the $\Gamma_n$ isomeric line. $\Gamma_\gamma^{(a)}(7p_{1/2}-7s;\omega) = 2.4\times10^{-6}$ eV.

\section{ Scan time estimate }
\label{scantime}

	For evaluation purposes, I consider a thought experiment involving scanning with a monochromatic light beam with a power of 1 mW. Such is the total power of all the teeth of the frequency comb in the  project \cite{zhang}.  Put $\omega$ = 10.75 eV --- resonance energy corresponding to the excitation of the $7p_{3/2}$ level in the final state. Let the target area  be $10^{-4}\text{ cm}^2$. Now the photon flux $J=5.47\times10^{18}\text{ cm}^{-2}\text{ c}^{-1}$, which is five orders of magnitude greater than in the frequency comb project. 

      By means of Eq. (\ref{si}) I find the cross section 
$\sigma_\gamma= 6.24\times10^{-21}\text{ cm}^2$ and estimate the rate of excitation of the isomer in the laser radiation field:
$p=\sigma_\gamma J = 0.034	\text{ c}^{-1} \ll \gamma$.
Thus, for the same $N_0$, in a time of 10$^{-5}$ s, $N_{is}\approx nN_0=5.4\times10^6$ isomeric atoms are formed, which provides $10^5$ times greater statistics as compared to the IC project. Therefore, this interval, that is, 10$^{-5}$ s, looks to be sufficient as the duration of the scanning pulse. If one puts the same time for checking if the condition of resonance is fulfilled, then each step will take $\tau_\text{sc}=2\times10^{-5}$ s.
      Furthermore, since the width of the $7p_{3/2}$ level is 
$\Gamma_{7p_{3/2}}=3.13\times10^{-8}$ eV, then  scanning an interval of 1 eV  will need $N_\text{ sc}=1\text{ eV} / \Gamma_{7p_{3/2}}=3.3\times10^7$ steps, that is $N_\text{sc}\tau_\text{sc}$ = 11 minutes. And for scanning the uncertainty interval of 0.024 eV half minute is enough. 

\begin{figure}[bth]
\includegraphics[width=0.8\textwidth]{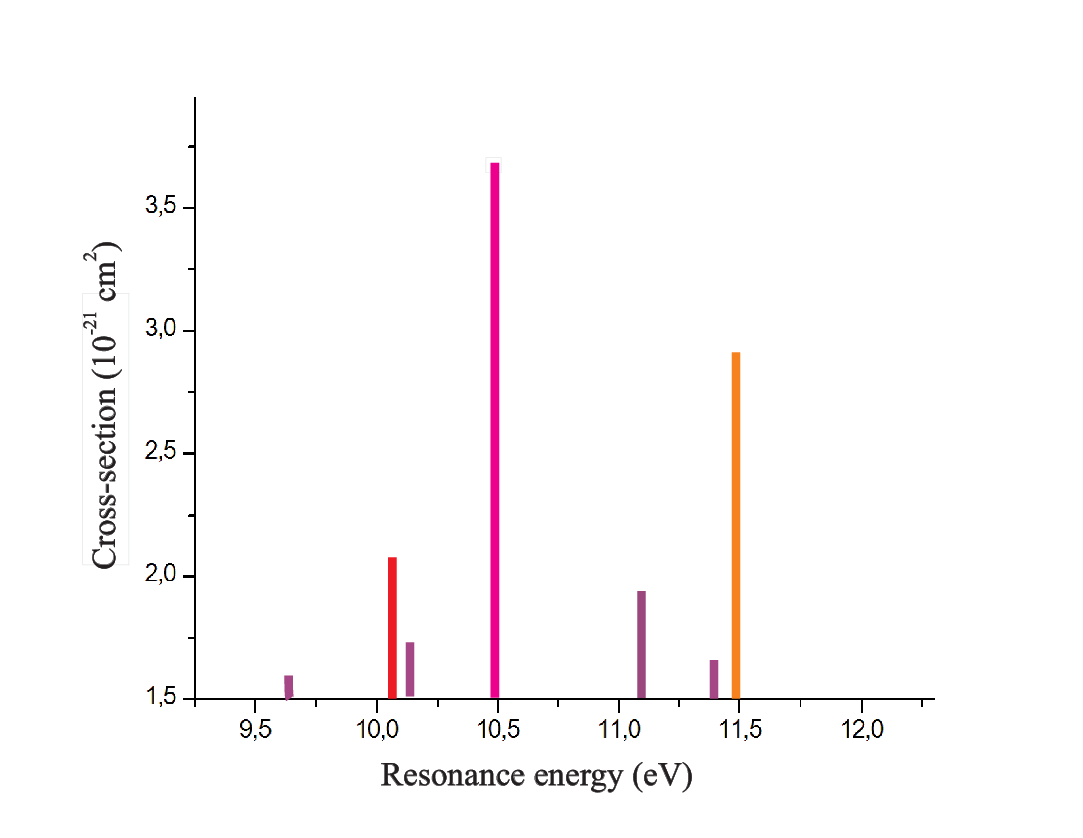}
\caption{\footnotesize Resonance landscape resulting from the fragmentation of the $7p$ levels 
according to \cite{nist}. }
\label{fThIres}
\end{figure}
Taking into account interelectronic interactions, the final $7p$ states are fragmented into several strong levels \cite{prc2008}, between which the resonance cross section found above is distributed. One can find the spectrum of the final states and the corresponding resonance cross sections using NIST data on the relative intensity of the Th I absorption lines from the ground state.
According to \cite{nist}, there is a strong component in the spectrum, corresponding to the transition from the ground state to a level with an energy of 2.1520074477 eV. Accordingly, this line forms the main resonance in scanning, achieved at a laser photon energy of $\omega$ = 10.490 eV. In Fig. \ref{fThIres},  the resonance cross sections are presented for the  strongest lines, appearing at energies $\omega$ = 11.476, 11.096, 10.067 eV and others. They can be used as reference points, the detection of which will help one to reliably identify these components in an experiment and, accordingly, determine the energy of the isomer according to (\ref{en}) with an accuracy corresponding to laser measurement methods. The cross-section scale is given under the assumption that the strongest line retains 1/3 of the total resonance strength of the $7p_{1/2}$- and $7p_{3/2}$ levels, as calculated within the framework of the Dirac-Fock method.

\section{ Discussion}
\label{conclusion}

To summarize, one can conclude  that when scanning  neutral atoms, one can take advantage of the broadening of the isomeric line due to IC by 9 orders of magnitude, up to $\sim10^{-10}$ eV. At the same time, this leads to a decrease in the photoexcitation cross section by the same factor. A frequency comb can help to reduce the scan time due to the large number of $\sim10^5$ teeth. A similar effect can be achieved with deliberately broadened spectral line of the scanning laser \cite{BCSH}.

The resonance mechanism of the line broadening, presented above, exploits big natural width of the atomic resonances involved. According  to this approach, the atom remains in an excited state with a good quality factor, as well as the nucleus does. This increases the line width to atomic values of $\sim10^{-8}$ eV. The isomer is excited due to the Feinberg---Migdal shaking mechanism. This involves the main smallness in the cross section of the process related with small probability of excitation of the isomer (Warsaw effect) $W = 4\times10^{-12}$  on the valence $7s$ electrons. This value coincides with that calculated in Ref. \cite{Letter}. The $W$ values increase on the inner shells, up to 10$^{-4}$ on the $1s$-shell \cite{npa,Letter,Kosel}. Another beautiful consequence of the Warsaw effect for the gyromagnetic ratio of the $^{229}$Th nucleus in the ground state was drawn attention to in Ref. \cite{shab}. The above estimates demonstrate robustness of the scheme. Thus, when comparing formulas (\ref{bn}) with (\ref{bn1}), the section $\sigma_\gamma^{(b.n.)}$ can be compared with the section $\tilde\sigma_\gamma^{(b.n.)}$ as quantities of the same order. However, the additional factor $fWQ\frac{\Gamma_{7p}}{\Gamma_\gamma^{(n)}}$= 92 speaks eloquently in favor of the two-photon resonance method. A similar  conclusion was also obtained in Ref. \cite{npa}.
	
      What is said above  fully applies to any other excitation schemes based on the absorption of two, three or more photons: excited electronic state, left after the nucleus is transferred to the isomeric state, radically contracts  scanning time needed. 
   For example, this moment was not taken into account in Ref. \cite{porsev}. It was shown already in Ref. \cite {yaf2015} that the cross-section radically increases if the electron shell is left in an excited state.  Multiplied by inconsistencies with multipolities of the comprising transitions,  this shortcoming gave rise to the thorium puzzle (e.g. \cite{np21} and Refs. therein). The above arguments sum up its solution.
   
\bigskip
   The author  would like to express his  gratitude to  P. G. Thirolf, L. F. Vitushkin and L. von der Wense for fruitful discussions.


\clearpage

\footnotesize
\begin{thebibliography}{99}

\bibitem{pik} E Peik, T Schumm, M S Safronova, A Palffy J. Weitenberg and P G Thirolf, Quantum Sci. Technol. 6 (2021) 034002.
\bibitem{kozclock} Steven A. King {\it et al.}. Nature, {\bf 611}, 43 (2022). 
\bibitem{dreif} Flambaum V 2006 Phys. Rev. Lett. 97 0 92502.
\bibitem{snowmass} D. Antypas {\it et al.}, arXiv:2203.14915 [hep-ex].
\bibitem{randevu} E. Peik, M. Okhapkin. {\it Nuclear clocks based on resonant 
excitation of $\gamma$-transitions}, C. R. Physique {\bf 16}, 516 (2015).

\bibitem{inter}  V V Flambaum and V A Dzuba. {\it  Search for variation of the fundamental constants in atomic, molecular, and nuclear spectra}, Can. J. Phys. {\bf 87}, 25 (2009).
\bibitem{BCSH} J. Tiedau, M. V. Okhapkin, K. Zhang {\it et al., Laser excitation of the Th-229 nucleus.} Phys. Rev. Lett., in print.

\bibitem{lars1} L. Von der Wense, B. Seiferle, M. Laatiaoui  {\it et al.} Direct detection of the $^{229}$Th nuclear clock transition, Nature, {\bf 47}, 533 (2016). 
\bibitem{prc2007} F. F. Karpeshin and M. B. Trzhaskovskaya. {\it Impact of the electron environment on the lifetime of the    $^{229m}$Th low-lying isomer.}     Phys. Rev. {\bf C76}, 054313 (2007).

\bibitem{isakov} N. Minkov, A. Palffy, Phys. Rev. Lett. 122, 162502 (2019).

\bibitem{zhang}  L. von der Wense  and  Z. Chuankun. Concepts for direct frequency-comb spectroscopy of 229mTh and an internal-conversion-based solid-state nuclear clock.  Eur. Phys. J. Ser. D {\bf 74},  146 (2020).

\bibitem{sandro} S. Kraemer, J. Moens, M. Athanasakis-Kaklamanakis {\it et al.} {\it Observation of the radiative decay of the $^{229}$Th nuclear clock isomer}.    Nature {\bf 617}, 706 (2023).
\bibitem{palfi1} Adriana  Palffy, Jorg Evers, and Christoph H. Keitel, Phys. Rev. C {\bf 77}, 044602 (2008).
\bibitem{palfi2} Adriana P?lffy, Oliver Buss, Axel Hoefer, and Hans A. Weidenm?ller, Phys. Rev. C {\bf 92}, 044619 (2015).
\bibitem{dzub1} A. Ya. Dzyublik,  G. Gosselin,  V. Meot and P. Morel, EPL, {\bf 102},  62001  (2013).
\bibitem{dzub2}  A. Ya. Dzyublik,  JETP Lett., {\bf 92}, 130 (2010).
\bibitem{wense} Lars von der Wense et al., EPJ A {\bf 56}, 176 (2020).

\bibitem{DF} D. F. Zaretsky {\it et al.} {\it Mesoatomic X-ray radiation of muons on    fragments  of instantaneous fission.} Yad. Fiz.  {\bf 29}, 306 (1979) [Sov. J. Nucl. Phys. {\bf 29}, 151 (1979)].

\bibitem{book} F.F.Karpeshin,  {\it Resonance Internal Conversion as the way of accelerating nuclear processes.}  Particles and Nuclei, {\bf 37}, 522 (2006). 

\bibitem{david} C.\,R\"{o}sel {\it et al.}    {\it Experimental evidence for muonic X-rays from fission fragments.}   Z. Phys. {\bf A345}, 425 (1993).

\bibitem{kabzon} B.A.Zon,  F.F.Karpeshin, {\it Acceleration of the decay of $^{235m}$U by laser-induced resonant     internal conversion.} Zh. Eksp. i  Teor. Fiz.,  {\bf 97}, 401 (1990) 
[Sov. Phys. --- JETP (USA), {\bf 70}, 224, 1990.]

\bibitem{krutov} V. A. Krutov. Ann. Phys. (Leipzig), {\bf 21}, 291 (1968); Pis'ma v ZhETF, {\bf 52}, 1176 (1990) [JETP Lett. {\bf 52}, 584 (1990)].

\bibitem{logan} D. Kekez {\it et al.}. Phys. Rev. Lett. {\bf 55}, 1366 (1985).

\bibitem{atalah}  F.\,F.\ Karpeshin,    M.R.Harston, F.Attallah,     J.F.Chemin,   J.N.Scheurer, I.M.Band, M.\,B.\ Trzhaskovskaya, {\it Subthreshold internal conversion to bound     states in highly-ionized $^{125}$Te ions}, Phys. Rev. {\bf C53} (1996) 1640.

\bibitem{morita} M. Morita,  Prog. Theor. Phys. {\bf 49}, 1574 (1973).
\bibitem{tkalya}  E.V.Tkalya. JETP Lett. {\bf 55}, 216 (1992); Nucl. Phys. A {\bf 539}, 209 (1992). 

\bibitem{batkin} I. S. Batkin. {\it Compton excitation of  the nuclear levels.} Yad. Fiz. {\bf 29}, 903 (1979)  [Sov. J. Nucl. Phys. {\bf 29}, 464 (1979)].
\bibitem{PL1}  F.F.Karpeshin,  I.M.Band, M.B.Trzhaskovskaya    and B.A.Zon,   
 {\it Study of \ $^{229}$Th through laser-induced resonance   internal conversion.} Phys. Lett. {\bf 282B}, 267 (1992). 

\bibitem{tom} T. T. Inamura and H. Haba, {\it Search for a ``3.5-eV isomer" in $^{229}$Th in a hollow-cathode electric discharge.} Phys. Rev. C {\bf 79}, 034313 (2009).

\bibitem{beck} {\it B. R. Beck et al.,} Phys. Rev. Lett., {\bf 98},  142501 (2007).

\bibitem{tai} P. V. Borisyuk, N. N. Kolachevsky, A. V. Taichenachev, E. V. Tkalya, I. Yu. Tolstikhina, and V. I. Yudin. {\it Excitation of the low-energy $^{229m}$Th isomer in the electron bridge pro-cess via the continuum}. Phys. Rev. C {\bf 100}, 044306 (2019).

\bibitem{dzub} A.Ya. Dzyublik. {\it Quasiclassical theory of 229mTh excitation by laser pulses via electron bridges}. Phys. Rev. C {\it 106}, 064608 (2022).

\bibitem{porsev} S. G. Porsev {\it et al.} {\it Excitation of the Isomeric $^{229m}$Th Nuclear State via an Electronic Bridge Process in $^{229}$Th$^+$}.  Phys. Rev. Lett.   {\bf 105},   182501 (2010).

\bibitem{prc2f}   F. F. Karpeshin, M. B. Trzhaskovskaya, {\it Bound internal conversion versus nu-clear excitation by electron transition: Revision of the theory of optical pumping of the $^{229m}$Th isomer.} Phys. Rev. {\bf C 95}, 034310 (2017).

\bibitem{muler} R.A. M{\"u}ller, A.V. Volotka, S. Fritzsche, A. Surzhykov, {\it Theoretical analysis of the electron bridge process in  $^{229}$Th$^{3+}$}. NIM B {\bf 408}, 84 (2017). 

\bibitem{Th35} Pavlo V. Bilous {\it et al.}, Phys. Rev. Lett. {\bf 124}, 192502 (2020).

\bibitem{2f} Haowei Xu, Hao Tang, Guoqing Wang, Changhao Li, Boning Li, Paola Cappellaro and Ju Li, Phys. Rev.  A {\bf 108}, L021502 (2023).
\bibitem{china} Lin Li {\it et al. Scheme for the excitation of thorium-229 nuclei based on electronic bridge excitation}, Nucl. Sci. Techn. (2023) 34:24; https://doi.org/10.1007/s41365-023-01169-4. 
\bibitem{2fchina} Neng-Qiang Cai, Guo-Qiang Zhang, Chang-Bo Fu and Yu-Gang Ma. {\it
Populating 229mTh via two-photon electronic bridge mechanism}, Nucl. Sci. Techn. (2021) 32:59; https://doi.org/10.1007/s41365-021-00900-3.
\bibitem{prev} L. von der Wense, B. Seiferle, S. Stellmer, J. Weitenberg, G. Kazakov,
A. Palffy, and P. G.  Thirolf. Phys. Rev. Lett. {\bf 119}, 132503 (2017).

\bibitem{Letter} F.F. Karpeshin, {\it Measurement of the Energy of the 8.3-eV $^{229}$Th Isomer 
Using the Photoelectric Effect.} JETP Letters, {\bf 118}, 548 (2023).

\bibitem{land} L. D. Landau and E. M. Lifshitz, {\it Course of Theoretical
Physics,} Vol. 3: {\it Quantum Mechanics: Non-Relativistic Theory} (Pergamon, New York, 1977).

\bibitem{npa} F. F. Karpeshin, I.M. Band, and M.B. Trzhaskovskaya, {\it  3.5-eV isomer of $^{229m}Th$: how it can be produced.} Nucl.  Phys. {\bf A654}, 579 (1999).

\bibitem{Arxiv}  F.F. Karpeshin  and L.F. Vitushkin. On the problems of creating a 
nuclear-optical frequency standard based on $^{229}$Th,  https://doi.org/10.48550/arXiv.2307.08711;  F.F.Karpeshin and M.B.Trzhaskovskaya,  Zh. Exp. Teor. Fiz. [JETP] {\bf 165}, 145 (2024).




\bibitem{beres} A.I.Akhiezer, V.B.Berestetskii. Quantum Electrodynamics. John Wiley \& Sons, New-York-London-Sydney, 1965.

\bibitem{Zyl} F.F.Karpeshin,  S.Wycech,  I.M.Band,     M.\,B.\ Trzhaskovskaya, M. Pf\"utzner and J.\.Zylicz,
{\it Rates of transitions between the  hyperfine-splitting components of the ground-state and the 3.5-eV isomer in $^{229}$Th$^{89+}$.}  Phys. Rev. {\bf C57}, 3085 (1998).

\bibitem {RAINE}  I. M. Band and  M. B. Trzhaskovskaya, Internal Conversion Coefficients for Low-Energy Nuclear Transitions, At. Data  Nucl. Data Tables {\bf 55}, 43 (1993).

\bibitem{prc2008} F.F.Karpeshin and M.B.Trzhaskovskaya,  {\it Impact of the electron environment on the lifetime of the     $^{229m}$Th low-lying isomer.}     Phys. Rev. {\bf C76}, 054313 (2008).

\bibitem{nist} A. Kramida and Yu. Ralchenko, J. Reader and NIST ASD Team (2022), NIST Atomic Spectra Database (ver. 5.10), https://physics.nist.gov/asd. National Institute of Standards and Technology, Gaithersburg, MD; DOI: https://doi.org/10.18434/T4W30F.

\bibitem {Kosel}  M. G. Kozlov, A. V. Oleynichenko, D. Budker, D. A. Glazov, Y. V. Lomachuk, V. M. Shabaev, A. V. Titov, I. I. Tupitsyn, and A. V. Volotka, 
Excitation of the $^{229}$Th nucleus by the hole in the inner electronic shells. Arxive: 2308.05173.

\bibitem {shab} Shabaev V.M., Glazov D. A., Ryzhkov A.M. {\it et al.}. Ground-state $g$ factor of highly charged $^{229}$Th ions: an access to the $M1$ transition probability between the isomeric and ground nuclear states. Phys. Rev. Lett. {\bf 128}, 043001 (2022).

\bibitem{yaf2015} F.F.Karpeshin and M.B.Trzhaskovskaya,  Yad. Fiz. {\bf  78}, 765 (2015) [Phys. At. Nucl. {\bf  78}, 715 (2015)].  
\bibitem{np21}  F.F.Karpeshin and M.B.Trzhaskovskaya,   A proposed solution for the lifetime puzzle of the $^{229m}$Th${^+}$ isomer,  Nucl. Phys. A{\bf1010},  122173 (2021).


\bibitem{arx}  F.F. Karpeshin  and L.F. Vitushkin. On the problems of creating a 
nuclear-optical frequency standard based on 229Th,  https://doi.org/10.48550/arXiv.2307.08711

\end{thebibliography}
\end{document}